\documentclass[12pt]{article}
\usepackage{enumitem}
\usepackage{fullpage}
\begin{document}
\title{A Lower Bound of $2^n$ Conditional Branches \\for Boolean Satisfiability on Post Machines}
\author{Samuel C. Hsieh\\Computer Science Department, Ball State University}
\maketitle
\begin{abstract}
We establish a lower bound of $2^n$ conditional branches for deciding the satisfiability of the conjunction of any two Boolean formulas from a set called a \textit{full representation} of Boolean functions of $n$ variables - a set containing a Boolean formula to represent each Boolean function of $n$ variables.  The contradiction proof first assumes that there exists a Post machine (Post's Formulation 1) that correctly decides the satisfiability of the conjunction of any two Boolean formulas from such a set by following an execution path that includes fewer than $2^n$ conditional branches. By using multiple runs of this Post machine, with one run for each Boolean function of $n$ variables, the proof derives a contradiction by showing that this Post machine is unable to correctly decide the satisfiability of the conjunction of at least one pair of Boolean formulas from a full representation of $n$-variable Boolean functions if the machine executes fewer than $2^n$ conditional branches. This lower bound of $2^n$ conditional branches holds for any full representation of Boolean functions of $n$ variables, even if a full representation consists solely of minimized Boolean formulas derived by a Boolean minimization method. We discuss why the lower bound fails to hold for satisfiability of certain restricted formulas, such as 2CNF satisfiability, XOR-SAT, and HORN-SAT. We also relate the lower bound to 3CNF satisfiability. The lower bound does not depend on sequentiality of access to the boxes in the symbol space and will hold even if a machine is capable of non-sequential access.
\end{abstract}
\section{Introduction}
The problem of deciding whether a Boolean formula is satisfiable is commonly known as the Boolean satisfiability problem. It was the first problem shown to be NP-complete [1]. This paper establishes a lower bound of $2^n$ conditional branches for deciding the satisfiability of the conjunction of any two Boolean formulas from a set called a \textit{full representation} of Boolean functions of $n$ variables - a set containing a Boolean formula to represent each Boolean function of $n$ variables.  The contradiction proof first assumes that there exists a Post machine (Post's Formulation 1) that correctly decides the satisfiability of the conjunction of any two Boolean formulas from such a set by following an execution path that includes fewer than $2^n$ conditional branches. By using multiple runs of this Post machine, with one run for each Boolean function of $n$ variables, the proof derives a contradiction by showing that this Post machine is unable to correctly decide the satisfiability of the conjunction of at least one pair of Boolean formulas from a  full representation of $n$-variable Boolean functions if the machine executes fewer than $2^n$ conditional branches.

We briefly summarize the remaining sections of this paper. The next section provides a brief overview of Boolean formulas and Post machines. As there are variations in the nomenclatures used in the literature, an overview of the related concepts and terminology as used in this paper seems appropriate. Section 3  introduces concepts related to  executing multiple runs of a Post machine and proves a few related lemmas. Section 4 proves the lower bound of $2^n$  conditional branches and shows that the lower bound applies to CNF satisfiability and, by duality, to DNF falsifiability. Section 5 first discusses why the lower bound fails to hold for satisfiability of certain restricted formulas such as 2CNF satisfiability, XOR-SAT and HORN-SAT, and the section then relates the lower bound to 3CNF satisfiability. Section 6 discusses  non-sequential access to the symbol space and similar proofs using Turing machines.
\section{Boolean Formulas and Post Machines}
Boolean formulas are widely known, e.g., [2,5]. As there are variations in the nomenclatures, we summarize the related concepts and terminology as used here.
\subsection{Boolean Formulas}
The set $B=\{true, false\}$ denotes the set of \textit{Boolean values}. A Boolean \textit{variable} has either $true$ or $false$ as its value. A function $f:B^n$ $\rightarrow$ $B$ is a Boolean function of $n$ variables. The expression $B^n \rightarrow B$ denotes the set of Boolean functions of $n$ variables. 

We may use a Boolean \textit{formula} to define or represent a Boolean function. A Boolean formula is composed of Boolean values, Boolean variables, and the Boolean operators $\wedge$ (for conjunction, i.e., AND), $\vee$ (for disjunction, i.e.,  OR) and $\overline{overbar}$ (for negation, i.e., NOT). A Boolean function can be represented by many different Boolean formulas.

A \textit{literal} is a Boolean variable or a logically negated variable. A Boolean formula in DNF  (\textit{Disjunctive Normal Form}), or a DNF formula, is an OR of DNF clauses, and a DNF clause is an AND of literals. For example, $( \overline{x_1} \wedge x_2 )\vee (x_1 \wedge \overline{x_2}) \vee (x_1 \wedge x_3)$  is a DNF formula. A DNF formula is a $k$DNF formula if each clause has $k$ literals. The example just given is a 2DNF formula. A Boolean formula in CNF (\textit{Conjunctive Normal Form}), or a CNF formula, is an AND of CNF clauses, and a CNF clause is an OR of literals. For example, $(x_1 \vee  x_2 ) \wedge (\overline{x_1} \vee \overline{x_2}) \wedge ( x_2 \vee x_3)$ is a CNF formula. A CNF formula is a $k$CNF formula if each clause has $k$ literals. The example just given is a 2CNF formula.

An \textit{assignment} to a set of Boolean variables assigns a Boolean value to each variable in the set. An assignment can be used to evaluate a Boolean formula or a Boolean function. If an assignment makes a formula or a function $true$,  the assignment is said to \textit{satisfy} the formula or the function and is called a \textit{satisfying} assignment; similarly, if an assignment makes a formula or a function $false$, the assignment is said to \textit{falsify} the formula or the function and is called a \textit{falsifying} assignment.

A Boolean formula is \textit{satisfiable} if it has a satisfying assignment; otherwise, the formula is \textit{unsatisfiable}. A Boolean formula is \textit{falsifiable} if it has a falsifying assignment; otherwise, the formula is \textit{unfalsifiable}.
\subsection{Post Machines}

Post's Formulation 1 [4], henceforth referred to as a \textit{Post machine}, is a very simple model of computation. A Post machine has access to a \textit{symbol space}, which is "a two-way infinite  sequence of boxes, ordinally similar to the series of integers $\cdots$, -3, -2, -1, 0, 1, 2, 3,$\cdots$" [4]. For convenience in our discussion, we will regard the series of integers as the \textit{addresses} of the boxes and usually refer to a specific box by its address, e.g., the box at address $x$ or the $x^{th}$ box. At any time, each box can be in one of two states: $marked$ or $blank$ (i.e., unmarked). A Post machine uses a \textit{read/write head} (called a "worker" in Post's paper [4]), henceforth referred to simply as a \textit{head}, to read (sense the state of) or to write (mark or unmark) the boxes in the symbol space, one box at a time. At any time, the head is positioned at some box. The address of the box where the head is positioned will be called the \textit{head position}. The box where the head is currently positioned is called the \textit{current box}. "One box is to be singled out and called the \textit{starting point}" [4], which will be called the \textit{initial head position} in this paper.

A Post machine executes a \textbf{fixed} finite program. The instructions (called "directions" in [4] ) in the program of a machine are "numbered 1, 2, 3,$\cdots$, n." In our discussion, we will consider the integers 1, 2, 3,$\cdots$, $n$ as the \textit{addresses} of the instructions and we will often refer to an individual instruction by its address, e.g., the instruction at address $x$. Execution of the program begins with the instruction at address 1. An instruction at any address $i$ can be of one of the following three types:
\begin{itemize}
\item[A)] Perform one of the following operations and then jump to the instruction at address $i_{next}$.
\begin{itemize}
	\item[a)] Mark the current box,\footnotemark[1]
	\item[b)] Unmark the current box (that is, make it blank),\footnotemark[1] 
\footnotetext[1]{A Post machine's fixed program is said to be applicable [4] to a general problem if, when applied to each instance of the general problem, the program never tries to mark an already marked box or to unmark an already blank box. Without loss of generality, we assume applicability of the fixed program of each machine in this paper.}
	\item[c)] Move the head right to the next box,
	\item[d)] Move the head left to the next box,
\end{itemize}
\item[B)] Sense the state of the current box and then, depending on whether the current box is marked or blank, jump to  the instruction at address $i_{marked}$ or $i_{blank}$ respectively,
\item[C)] Stop, that is, halt the machine.
\end{itemize}

Clearly, executing an instruction of type B, henceforth referred to as a \textit{conditional branch}, makes a binary decision: if the current box is marked, then jump to the instruction at address $i_{marked}$ else jump to the instruction at address $i_{blank}$. On the other hand, an instruction of type A specifies the instruction at address $i_{next}$ as the instruction to execute next and does not involve a decision.  

A problem to be solved by a Post machine is to be given and its answer is to be received as configurations of a finite number of marked boxes in the symbol space by an "outside agency" [4]. Those boxes not used to represent an input are initially blank. Just as binary strings can be used to represent various kinds of information, sequences of $blank$ and $marked$ boxes can be similarly used.  For the Boolean satisfiability problem, such sequences may, for example, be used to represent the symbols of an appropriate alphabet from which strings can be composed to represent formulas and other information. Such an alphabet may, for example, include symbols for the Boolean operators, Boolean values, and parentheses, as well as symbols to form strings to represent identifiers. An unlimited number of identifiers, values, and formulas can be represented as strings formed from a fixed finite alphabet. 

For a decision problem, such as the Boolean satisfiability problem, an answer is either yes or no. We will say that a Post machine \textbf{accepts} its input if the machine halts with the answer yes and that a machine \textbf{rejects} its input if the machine halts with the answer no.

\section{Running a Post Machine on a Bipartite Input}
First, we will define several related terms. 
\begin{flushleft}
\textbf{Definition 1}. A \textit{partition} is a subset of a symbol space. 
\end{flushleft}

We will run a Post machine to decide the satisfiability of the conjunction of two Boolean formulas. Hence, an input will consist of two parts, one for each conjunct. The two parts of an input will be provided in two \textbf{disjoint} partitions. It is trivial to divide the boxes of a symbol space into two disjoint partitions. As an example, one partition may consist of the boxes with addresses greater than some arbitrary integer $x$, with the other partition consisting of those boxes with addresses $\leq x$. As another example, one partition may consist of those boxes with addresses that are even numbers, with the other partition consisting of those with odd addresses.
\begin{flushleft}
\textbf{Definition 2}.  A \textit{bipartite} input consists of two parts that are provided in a symbol space divided into two disjoint partitions, with each part of the input placed in a separate partition.
\end{flushleft}

A bipartite input will be denoted as an ordered pair $(first,$ $second)$, where $first$ and $second$ denote the first and the second parts of the input. The two partitions where $first$ and $second$ are provided will be called the first and the second partition respectively.

A convention for dealing with the following issues for a Post machine that takes a bipartite input is called a \textit{symbol space convention}: where to position the head initially, how the symbol space is divided into disjoint partitions, and where to place each part of a bipartite input in its partition. 

\begin{flushleft}
\textbf{Definition 3}. A \textit{symbol space convention} refers to  a precise specification of the following:
\end{flushleft}
\begin{itemize}
\item[a)] a specific address as the initial head position, i.e., the address of the box "singled out and called the starting point" [4], and  
\item[b)] division of the symbol space into two disjoint partitions, and  
\item[c)] location of each part of a bipartite input in its corresponding partition.
\end{itemize}

As an example, a symbol space convention  may, quite arbitrarily, specify that a) the head is to be initially positioned at the address 0, b) the first partition consists of those boxes at addresses $<$ 0, and the second partition consists of those boxes at addresses $\geq$ 0, and c) the first part of a bipartite input is to be located in the boxes at the addresses $\cdots,$  $-3,$ $-2,$ $-1$, with the rightmost box of the first part of the input located at the address -1, and the second part of a bipartite input is to be located in the boxes at the addresses $0,$ $1,$ $2,$ $\cdots$, with the leftmost box of the second part of the input located at the address 0. 	
\begin{flushleft}
\textbf{Definition 4}. An execution of a Post machine M on a  bipartite input $(first, second)$ is called a \textit{run} of M and is denoted by M$(first,$ $second)$. An identical symbol space  convention is adopted for all runs of a given Post machine.
\end{flushleft}

A run of a Post machine solves an \textbf{instance} of the general problem that the Post machine is intended to solve. For example, if a machine M solves the Boolean satisfiability problem, then a run of M decides whether a specific Boolean formula is satisfiable.

A symbol space convention for a Post machine is analogous to an input convention assumed by a computer program: where the program's input is provided and how the input data are organized. That an identical symbol space convention is adopted for all runs of a given Post machine is analogous to that an identical input convention is assumed by all executions of a given program. Since the same symbol space convention is adopted for all runs of a given Post machine, each run executes with the same initial head position and with the symbol space divided into partitions in the same way. Besides, if two runs are given the same first (or second) part of a bipartite input, the initial content of the first (or second, respectively) partition for one run will be identical to that for the other run: that is, for every address $x$ in the first (or second, respectively) partition, the $x^{th}$ box for one run will have the same initial state (possibly $blank$) as the $x^{th}$ box for the other run.

To illustrate bipartite inputs for multiple runs of a Post machine, let us look at an example. Let M$(first_1,$ $second_1)$, M$(first_2,$ $second_2)$, M$(first_1,$ $second_2)$, M$(first_2,$ $second_1)$ be four runs of a machine M, where $first_1=$ $mbm$, $second_1=$ $mbbm$,  $first_2=$ $mmbm$, and $second_2=$ $mbmm$, where $b$ denotes a blank box and $m$ denotes a marked box. Suppose the adopted symbol space convention specifies, rather arbitrarily, that
\begin{itemize}
\item the first partition consists of those boxs at addresses $< 15$ and the second partition consists of those at addresses $\geq 15$, and
\item the rightmost box of the first part of each bipartite input is located at the address 14 and the leftmost box of the second part of each bipartite input is located at the address 15.
\end{itemize}

\noindent
The following table (Table 1) shows how the four bipartite inputs are provided for the four runs.

\begin{center}
\begin{tabular}{l|c|c|c|c|c|c|c|c|c|c|c|c|}
\cline{2-13}
& \multicolumn{6}{ |c| }{First partition}& \multicolumn{6}{ c|}{Second partition}\\ 
\hline
\multicolumn{1}{ |c| }{Addresses}&09&10&11&12&13&14&15&16&17&18&19&20 \\ 
\noalign{\hrule height 2pt}
\multicolumn{1}{ |c| }{M$(first_1,$ $second_1)$}&$b$ &$b$ &$b$ &\textbf{\textit{m}}&\textbf{\textit{b}}&\textbf{\textit{m}}&\textbf{\textit{m}}&\textbf{\textit{b}}&\textbf{\textit{b}}&\textbf{\textit{m}}&$b$ &$b$  \\ 
\hline
\multicolumn{1}{ |c| }{M$(first_2,$ $second_2)$}&$b$ &$b$ &\textbf{\textit{m}}&\textbf{\textit{m}}&\textbf{\textit{b}}&\textbf{\textit{m}}&\textbf{\textit{m}}&\textbf{\textit{b}}&\textbf{\textit{m}}&\textbf{\textit{m}} &$b$ &$b$ \\ 
\hline
\multicolumn{1}{ |c| }{M$(first_1,$ $second_2)$}&$b$ &$b$ &$b$ &\textbf{\textit{m}}&\textbf{\textit{b}}&\textbf{\textit{m}}&\textbf{\textit{m}}&\textbf{\textit{b}}&\textbf{\textit{m}}&\textbf{\textit{m}}&$b$ &$b$  \\ 
\hline
\multicolumn{1}{ |c| }{M$(first_2,$ $second_1)$}&$b$ &$b$ &\textbf{\textit{m}}&\textbf{\textit{m}}&\textbf{\textit{b}}&\textbf{\textit{m}}&\textbf{\textit{m}}&\textbf{\textit{b}}&\textbf{\textit{b}}&\textbf{\textit{m}} &$b$ &$b$ \\ 
\hline
\multicolumn{13}{ c }{Table 1. Bipartite Inputs for Multiple Runs} \\
\end{tabular}
\end{center}
\begin{flushleft}
Since the same symbol space convention is adopted for these runs, the first partition for M$(first_1,$ $second_1)$ is identical to that for M$(first_1,$ $second_2)$ because the two runs have the same first part $first_1$ in their bipartite inputs. Similarly, the second partition for M$(first_2,$ $second_2)$ is identical to that for M$(first_1,$ $second_2)$. So is the first partition for M$(first_2,$ $second_2)$ to that for M$(first_2,$ $second_1)$, and so too is the second partition for M$(first_1,$ $second_1)$ to that for M$(first_2,$ $second_1)$.
\end{flushleft} 
\begin{flushleft}
\textbf{Definition 5}. An \textit{execution path}, or simply a \textit{path}, is a sequence of instructions that a Post machine may execute, beginning with the instruction at the address 1. A path is either \textit{terminated} or \textit{open}: a terminated path ends with the instruction Stop, and an open path ends with a type-A instruction or a conditional branch . A Post machine that serially executes the entire sequence of instructions of a path is said to \textit{follow} the path. The first instruction that a machine executes after following an open path P or after executing an instruction d is said to \textit{immediately succeed} the path P or the instruction d. An instruction  that immediately succeeds a path P or an instruction d  is called an \textit{immediate successor instruction} of P or of d.  
\end{flushleft}

The immediate successor instruction of a conditional branch (i.e., type-B instruction) depends on whether the current box is blank or marked, but the immediate successor instruction of a type-A instruction is \textbf{fixed} for the instruction and does not depend on the state of the current box. A terminated path has no successor instruction.

\begin{flushleft}
\textbf{Lemma 1}. For any Post machine and for any integer $m$ $\geq$ 0, the sum of the following two numbers is no more than $2^m$.
\end{flushleft}

\begin{itemize}
\item[a)]the number of distinct open paths with each containing $m+1$ conditional branches and ending with a conditional branch, and  
\item[b)] the number of distinct terminated paths with each containing $m$ or fewer conditional branches.
\end{itemize} 
\textbf{Proof}. We define a term to be used later in the proof: a \textit{line} is a sequence of instructions that a Post machine may execute and that consists of 0 or more type-A instructions followed either by a conditional branch or by the  instruction Stop. Since each type-A instruction specifies a \textbf{fixed} immediate successor instruction, if the first instruction of a line is executed, all instructions of the line will be executed serially. Every instruction d in any Post machine's fixed program begins either one line or no line. This is because  if d is the Stop instruction or a conditional branch, d alone is a line; if d is a type-A instruction, in any line that begins with d, every type-A instruction has a fixed immediate successor, and hence there is only one possible sequence of  execution and no alternative is possible until the end of the line, which is either a conditional branch or the Stop instruction.  It is possible for a type-A instruction to begin no line, e.g., when the instruction is in a loop that has no exit. Hence, every instruction in a Post machine's fixed program begins no more than one line.

We will prove the lemma by induction. The instruction at the address 1 begins no more than one line, which ends with either the Stop instruction or a conditional branch. In the former case, the line is one  ($=2^0$) terminated path containing no conditional branch; in the latter case, there is one open path containing 1 conditional branch. Hence the lemma holds for $m=0$. Suppose that there are $t$ distinct terminated paths with each containing $i$ or fewer conditional branches,  that there are $p$ open paths with each containing $i+1$ conditional branches and ending with a conditional branch, and that $t+p$ $\leq$ $2^i$ (inductive hypothesis). Since each of the $p$ open paths ends with a conditional branch, each open path has no more than two alternative immediate successor instructions, each of which begins no more than one line. If an alternative immediate successor of an open path begins a line ending with the Stop instruction, the open path appended with the line forms a terminated path containing $i+1$ conditional branches; on the other hand, if an alternative immediate successor of an open path begins a line ending with a conditional branch, the open path appended with the line forms an open path containing $i+2$ conditional branches. Thus, the p open paths and the lines that their alternative immediate successors begin can form no more than $2p$ distinct paths, some of which will remain open with each containing $i+2$ conditional branches, but the others will become terminated, with each containing $i+1$ conditional branches. No new paths can be formed from the $t$ terminated paths with each containing $i$ or fewer conditional branches since terminated paths have no successor instruction. Hence the sum of the number of distinct open paths with each containing $i+2$ conditional branches and ending with a conditional branch, and the number of terminated paths with each containing $i+1$ or fewer conditional branches, is no more than  $t+2p$, which is no more than $2^{i+1}$ since by the inductive hypothesis $t+p$ $\leq$ $2^i$.
\textbf{Q.E.D.}   
\begin{flushleft}
\textbf{Lemma 2}. If two runs M$(first_1,$ $second_1)$ and M$(first_2,$ $second_2)$ of a Post machine M follow a terminated path P, then the run M$(first_1,$ $second_2)$ must follow the same terminated path P.
\end{flushleft}
\textbf{Proof}. We will prove the lemma by contradiction. Let the path P, which the two runs M$(first_1,$ $second_1)$ and M$(first_2,$ $second_2)$ follow, be the sequence of instructions $P_1$ $P_2$ $\cdots$ $P_p$ , let the run M$(first_1,$ $second_2)$ follow the path Q, and let Q be the sequence of instructions $Q_1$ $Q_2$ $\cdots$ $Q_q$.  Assume that Q is different from P. The rest of the proof will derive a contradiction to this assumption. 

Since Q is different from P, there is at least one integer $i$ such that the instruction $Q_i$ is different from the instruction $P_i$. Of such integers, there must a least one. Let $m$ be the least such integer. Since $m$ is the smallest integer such that $Q_m$ is different from $P_m$, the path $P_1$ $\cdots$ $P_{m-1}$ is identical to the path $Q_1$ $\cdots$ $Q_{m-1}$. Since all runs begin with the same instruction (the instruction at address 1), $m$ $\geq$ 2. Since $P_m$ and $Q_m$ are different instructions, the instruction $P_{m-1}$, which is the same instruction as $Q_{m-1}$, has two different immediate successor instructions and therefore must be a conditional branch. Let T be the point in P and in Q between the instruction $P_{m-2}$ and the instruction $P_{m-1}$ (i.e., between $Q_{m-2}$ and $Q_{m-1}$). Let the three runs execute to the point T, where each run has completed the common sequence of instructions $P_1$ $\cdots$ $P_{m-2}$ (i.e., $Q_1$ $\cdots$ $Q_{m-2}$) but has not executed the instruction $P_{m-1}$ (i.e., $Q_{m-1}$) yet. Since each run begins with the head located at the same initial position (at the box that is "singled out and called the starting point"[4]), and in following the common path $P_1$ $\cdots$ $P_{m-2}$ to the point T, each of the three runs performs an identical sequence of  operations to move the head, at the point T each run has its head  positioned at a common address. Let $x$ be the address of this common head position. Consider the $x^{th}$ box for the run M$(first_1,$ $second_2)$: the box either has been written (i.e., marked or unmarked) by an instruction in the path $P_1$ $\cdots$ $P_{m-2}$, or it has not. In either case, $P_m$ and $Q_m$ can be shown to be the same instruction, as detailed below. 

\begin{itemize}
\item[A)]Suppose the $x^{th}$ box for M$(first_1,$ $second_2)$ has been written by an instruction in the path $P_1$ $\cdots$ $P_{m-2}$. Since all three runs follow the common path $P_1$ $\cdots$ $P_{m-2}$ to the point T,  all three runs perform an identical sequence of operations, including operations to mark or unmark the boxes, as they follow the common path to the point T.  Hence, at the point T the $x^{th}$ box for M$(first_1,$ $second_2)$ and the corresponding box for each of the other two runs must be in the same state, which will cause $P_{m-1}$ and $Q_{m-1}$, which is the same conditional branch as $P_{m-1}$, to select the same immediate successor instruction. That is, $P_m$ is the same instruction as $Q_m$.
\item[B)]Suppose the $x^{th}$ box for M$(first_1,$ $second_2)$ has \textbf{not} been written by an instruction in the path $P_1$ $\cdots$ $P_{m-2}$. The address $x$ is either in the first partition or in the second. In either case, $P_m$ and $Q_m$ can be shown to be the same instruction, as detailed below.
\begin{itemize}
\item[B.1)]Suppose $x$ is in the first partition. Since M$(first_1,$ $second_2)$ and M$(first_1,$ $second_1)$ have an identical first part in their bipartite inputs, both runs are given identical initial content in their first partition. Since the $x^{th}$ box has not been written along the path $P_1$ $\cdots$ $P_{m-2}$, at the point T the $x^{th}$ box for M$(first_1,$ $second_2)$ and the corresponding box for M$(first_1,$ $second_1)$ must remain in their common initial state. Hence, at the point T, $P_{m-1}$ and $Q_{m-1}$, which is the same conditional branch as $P_{m-1}$, will select the same immediate successor instruction. That is, $P_m$ is the same instruction as $Q_m$.
\item[B.2]Suppose $x$ is in the second partition. Similarly to case B.1 the $x^{th}$ box for M$(first_1,$ $second_2)$ and the corresponding box for M$(first_2,$ $second_2)$ can be shown to be in the same state at the point T, and hence $P_{m-1}$ and $Q_{m-1}$, which is the same conditional branch as $P_{m-1}$, will select the same immediate successor instruction. That is, $P_m$ is the same instruction as $Q_m$.
\end{itemize} 
\end{itemize}
\begin{flushleft}
Thus, there does not exist an integer $m$ such that $Q_m$ is different from $P_m$. In other words, the path P and the path Q are identical. \textbf{Q.E.D.}
\end{flushleft}

It is interesting to note that Lemma 2 holds no matter which box is singled out to be the starting point, no matter how the symbol space is divided into disjoint partitions, and no matter where each part of a bipartite input is placed in its corresponding partition. In short, Lemma 2 holds no matter what symbol space convention is adopted, as long as an identical symbol space convention is adopted for the runs involved.

\section{A Lower Bound for Satisfiability}

We now establish a worst-case lower bound on the number of conditional branches required to decide Boolean satisfiability on a Post machine.

\begin{flushleft}
\textbf{Definition 6}. Let $x_1,$ $x_2$ $\cdots$ $x_n$ be the  variables of which the members of the set $B^n$ $\rightarrow$ $B$ are functions. A Boolean formula that  \textit{represents} or \textit{defines} a Boolean function $f:B^n$ $\rightarrow$ $B$ is a formula $\phi_f$ of the variables $x_1,$ $x_2$ $\cdots$ $x_n$ such that, for every assignment to the variables $x_1,$ $x_2$ $\cdots$ $x_n$, the formula $\phi_f$ evaluates to the same value as what the function $f$ evaluates to. 
\end{flushleft}

A Boolean formula that represents a function $f$ will be denoted by $\phi_f$ here. However, the symbol $\phi$ without a subscript, or with a numerical subscript, such as $\phi_3$, will denote a Boolean formula without indicating the specific function that it represents.

\begin{flushleft}
\textbf{Definition 7}. Let $x_1,$ $x_2$ $\cdots$ $x_n$ be the  variables of which the members of the set $B^n$ $\rightarrow$ $B$ are functions. A \textit{full representation} of the set $B^n$ $\rightarrow$ $B$ is a set $E$ of Boolean formulas of the variables $x_1,$ $x_2$ $\cdots$ $x_n$ such that  every function $f:B^n$ $\rightarrow$ $B$ is represented by a formula $\phi_f$ $\in$ $E$. The set $E$ is said to \textit{fully represent} the set $B^n$ $\rightarrow$ $B$.
\end{flushleft}

\begin{flushleft}
\textbf{Definition 8}. The \textit{logical negation} of a function $g:B^n$ $\rightarrow$ $B$ is a function $\overline{g}:B^n$ $\rightarrow$ $B$ such that, on every assignment to the variables $x_1,$ $x_2$ $\cdots$ $x_n$,
\end{flushleft}
\begin{center}
$\overline{g}(x_1,$ $x_2$ $\cdots$ $x_n)$ = $\overline{g(x_1, x_2 \cdots x_n)}$.
\end{center}

The logical negation of a function $g$ is denoted by $\overline{g}$. For any function $g:B^n$ $\rightarrow$ $B$ and for every assignment, $g$ and $\overline{g}$ must evaluate to different values: one of them must evaluate to $false$ and the other to $true$.

\begin{flushleft}

\textbf{Definition 9}. A run M$(\phi_1,$ $\phi_2)$ is said to \textit{decide the satisfiability of $\phi_1$ $\wedge$ $\phi_2$}, or to \textit{decide whether $\phi_1$ $\wedge$ $\phi_2$ is satisfiable}, if and only if the run accepts its input (i.e., halts with the answer yes) if $\phi_1$ $\wedge$ $\phi_2$ is satisfiable and rejects its input (i.e., halts with the answer no) otherwise. A run M$(\phi_1,$ $\phi_2)$ is said to \textit{decide the falsifiability of $\phi_1$ $\vee$ $\phi_2$}, or to  \textit{decide whether $\phi_1$ $\vee$ $\phi_2$ is falsifiable}, if and only if the run accepts its input if $\phi_1$ $\vee$ $\phi_2$ is falsifiable and rejects its input otherwise.
\end{flushleft}

\begin{flushleft}
\textbf{Theorem 1}. Let $E$ be a full representation of the set  $B^n$ $\rightarrow$ $B$. There \textbf{does not exist} a Post machine M such that, for every pair of formulas $\phi_1,$ $\phi_2$ $\in$ $E$, M$(\phi_1,$ $\phi_2)$ correctly decides  whether $\phi_1$ $\wedge$ $\phi_2$ is satisfiable by by following a terminated path that includes fewer than $2^n$ conditional branches.
\end{flushleft} 
\textbf{Proof}. We will prove the theorem by contradiction. We first assume that there exists a Post machine M such that, for every pair of formulas $\phi_1,$ $\phi_2$ $\in$ $E$, M$(\phi_1,$ $\phi_2)$ correctly decides the satisfiability of $\phi_1$ $\wedge$ $\phi_2$ by following a terminated path that includes fewer than $2^n$ conditional branches. In other words, M$(\phi_1,$ $\phi_2)$ will accept the input if $\phi_1$ $\wedge$ $\phi_2$ is satisfiable and reject the input otherwise, and M$(\phi_1,$ $\phi_2)$ will do so by following a terminated path that includes fewer than $2^n$ conditional branches. The rest of this proof will derive a contradiction to this assumption.

Since $E$ fully represents the set $B^n$ $\rightarrow$ $B$, every function $f:B^n$ $\rightarrow$ $B$ and its logical negation $\overline{f}:B^n$ $\rightarrow$ $B$ are represented by some Boolean formulas $\phi_f,$ $\phi_{\overline{f}}$ $\in$ $E$. Let S be a set containing, for each distinct function $f:B^n$ $\rightarrow$ $B$, one run of M with  $(\phi_f,$ $\phi_{\overline{f}})$ as its bipartite input.  In other words, for each function $f:B^n$ $\rightarrow$ $B$, S contains the run M$(\phi_f,$ $\phi_{\overline{f}})$, which is to decide whether the formula  $\phi_f$ $\wedge$ $\phi_{\overline{f}}$ is satisfiable. Since there are $F=$ $2^{2^n}$ distinct functions in the set $B^n$ $\rightarrow$ $B$, the set S has $F$ runs of the machine M. 

The set S may seem expensive to implement in terms of computing resources. However, S will only be used to prove that logically the Turing machine M does not exist. An actual implementation of S is not needed.

Since, for every function $f:B^n$ $\rightarrow$ $B$ and for every assignment, either the function $f$ or its logical negation $\overline{f}$ evaluates to $false$, and since $\phi_f,$ $\phi_{\overline{f}}$ $\in$ $E$ represent $f$ and $\overline{f}$, for every assignment either $\phi_f$ or $\phi_{\overline{f}}$ evaluates $false$. Therefore, the formula $\phi_f$ $\wedge$ $\phi_{\overline{f}}$ is $false$ for every assignment and, thus,	 is not satisfiable. Hence, every run in the set S must eventually reject its input. By our assumption on M, every run in S must follow a terminated path that includes $2^n-1$ or fewer conditional branches and reject its input. 

By Lemma 1, there are no more than $2^m$ terminated paths with each including $m$ or fewer conditional branches. Since each run in S follows a terminated path that includes $2^n-1$ or fewer conditional branches, by Lemma 1 there are no more than the following number of terminated paths that the runs in S may follow.
\begin{center}
$2^{(2^n-1)} = 2^{2^n} 2^{-1} = F 2^{-1}  = F/2$
\end{center}

To summarize, each of the $F =$ $2^{2^n}$ runs in the set S follows a terminated path that includes $2^n-1$ or fewer conditional branches to reject its input, but there are no more than $F/2$ such paths. Therefore, there is at least one such path that multiple runs in S follow. Let P be a path that multiple runs in S follow and let M$(\phi_g,$ $\phi_{\overline{g}})$  and M$(\phi_h,$ $\phi_{\overline{h}})$ be two runs in S that follow the path P. Since S contains one run of M for each distinct Boolean function of $n$ variables, $g$ and $h$ must be different functions.  Since, as discussed previously, all runs in S must reject their inputs,  both M$(\phi_g,$ $\phi_{\overline{g}})$  and M$(\phi_h,$ $\phi_{\overline{h}})$ must reject their inputs. By Lemma 2,  two other runs, M$(\phi_g,$ $\phi_{\overline{h}})$  and M$(\phi_h,$ $\phi_{\overline{g}})$, which are not in S, must follow the same path P and \textbf{reject} their inputs, as the two runs M$(\phi_g,$ $\phi_{\overline{g}})$  and M$(\phi_h,$ $\phi_{\overline{h}})$ do. 

Now let us derive a contradiction to the assumption that the Post machine M exists. Since $g$ and $h$ are different Boolean functions, there exists an assignment $s$ that makes $g$ and $h$ evaluate to different values. Hence, the assignment $s$ will make $g$ and $\overline{h}$ evaluate to the same value. If both $g$ and $\overline{h}$ evaluate to $true$ on the assignment $s$, so will both of the formulas $\phi_g$ and $\phi_{\overline{h}}$, since $\phi_g,$ and $\phi_{\overline{h}}$ represent $g$ and $\overline{h}$. Thus, $\phi_g$ $\wedge$ $\phi_{\overline{h}}$ is satisfiable. On the other hand, if $g$ and $\overline{h}$ evaluate to $false$ on the assignment $s$, then $h$ and $\overline{g}$ will evaluate to $true$ on the assignment $s$ and so will the formulas $\phi_h$ and $\phi_{\overline{g}}$, since $\phi_h$ and $\phi_{\overline{g}}$  represent $h$ and $\overline{g}$. Thus, $\phi_h$ $\wedge$ $\phi_{\overline{g}}$ is satisfiable. Therefore, at least one of the two formulas $\phi_g$ $\wedge$ $\phi_{\overline{h}}$  and $\phi_h$ $\wedge$ $\phi_{\overline{g}}$ is satisfiable and, thereby, at least one of the two runs M$(\phi_g,$ $\phi_{\overline{h}})$  and M$(\phi_h,$ $\phi_{\overline{g}})$ \textbf{should accept} its input. However, as discussed previously, by Lemma 2 both M$(\phi_g,$ $\phi_{\overline{h}})$  and M$(\phi_h,$ $\phi_{\overline{g}})$ \textbf{reject} their inputs. That is, by Lemma 2, at least one of the two runs M$(\phi_g,$ $\phi_{\overline{h}})$  and M$(\phi_h,$ $\phi_{\overline{g}})$ \textbf{incorrectly} rejects its input. This contradicts our assumption that the machine M exists such that, for every pair of formulas $\phi_1,$ $\phi_2$ $\in$ $E$, M$(\phi_1,$ $\phi_2)$ \textbf{correctly} decides the satisfiability of $\phi_1$ $\wedge$ $\phi_2$ by following a terminated path that includes fewer than $2^n$ conditional branches. \textbf{Q.E.D.}

\null

By Theorem 1, for any Post machine M, there is at least one pair of formulas $\phi_1$ and $\phi_2$ in any full representation of $B^n$ $\rightarrow$ $B$ such that M$(\phi_1,$ $\phi_2)$ cannot correctly decide the satisfiability of $\phi_1$ $\wedge$ $\phi_2$  by executing fewer than $2^n$ conditional branches. In other words, $2^n$ is a lower bound on the number of conditional branches needed.

Like Lemma 2, Theorem 1 holds regardless of the initial head position, the way the symbol space is divided into disjoint partitions, and the location of each part of a bipartite input in its corresponding partition. In short, the theorem holds no matter what symbol space convention is adopted, as long as an identical symbol space convention is adopted for the runs involved. Besides, it should be noted that the proof for Theorem 1 does not rely on a specific representation of Boolean functions. As a result, the lower bound applies to the problem of deciding whether the conjunction of a pair of $n$-variable Boolean functions has a satisfying assignment, even if the two conjuncts are represented in the input as some expressions other than the form of Boolean formulas introduced in Section 2.1. 

Since there are many different Boolean formulas that represent a given Boolean function, there are many full representations of the set $B^n$ $\rightarrow$ $B$. It is interesting to note that Theorem 1  holds for any full representation $E$, even if $E$ consists solely of minimized Boolean formulas that are derived by a Boolean minimization method.

Since the set $B^n$ $\rightarrow$ $B$ can be fully represented by a set of CNF formulas, the lower bound holds even if the conjuncts $\phi_1$ and $\phi_2$ are limited to CNF formulas. 
\begin{flushleft}
\textbf{Corollary 1.1}. Let $E$ be a set of CNF formulas that fully represents  $B^n$ $\rightarrow$ $B$. There \textbf{does not exist} a Post machine M such that, for every pair of formulas $\phi_1,$ $\phi_2$ $\in$ $E$, M$(\phi_1,$ $\phi_2)$ correctly decides  whether the CNF formula $\phi_1$ $\wedge$ $\phi_2$ is satisfiable by following a terminated path that includes fewer than $2^n$ conditional branches.
\end{flushleft}

By duality, Corollary 1.2 follows from Theorem 1:
\begin{flushleft}
\textbf{Corollary 1.2}. Let $E$ be a full representation of $B^n$ $\rightarrow$ $B$. There \textbf{does not exist} a Post machine M  such that, for every pair of formulas $\phi_1,$ $\phi_2$ $\in$ $E$, M$(\phi_1,$ $\phi_2)$ correctly decides  whether $\phi_1$ $\vee$ $\phi_2$ is falsifiable by following a terminated path that includes fewer than $2^n$ conditional branches.
\end{flushleft}

By duality, Corollary 1.3 follows from Corollary 1.1.
\begin{flushleft}
\textbf{Corollary 1.3}. Let $E$ be a set of DNF formulas that fully represents $B^n$ $\rightarrow$ $B$. There \textbf{does not exist} a Post machine M such that, for every pair of formulas $\phi_1,$ $\phi_2$ $\in$ $E$, M$(\phi_1,$ $\phi_2)$ correctly decides  whether the DNF formula $\phi_1$ $\vee$ $\phi_2$ is falsifiable by following a terminated path that includes fewer than $2^n$ conditional branches.
\end{flushleft}
\section{Restricted Formulas}

Theorem 1 requires that the two conjuncts $\phi_1$ and $\phi_2$ be members of a full representation of $B^n$ $\rightarrow$ $B$. Since the following widely known sets of restricted formulas of $n$ variables do not fully represent $B^n$ $\rightarrow$ $B$,  Theorems 1 does not apply if the two conjuncts are limited to $n$-variable formulas from any of these sets: XOR-SAT, HORN-SAT, 2CNF, and 3CNF. Polynomial-time algorithms to decide 2CNF satisfiability,  XOR-SAT, and HORN-SAT are known. The next theorem establishes a lower bound of $2^n$ conditional branches for 3CNF satisfiability.

\begin{flushleft}
\textbf{Definition 10}. Let $E_1$ and $E_2$ be sets of Boolean formulas. A \textit{satisfiability-preserving} mapping from $E_1$ to $E_2$ is a function $t:E_1$ $\rightarrow$ $E_2$ such that, for every formula $\phi$ $\in$ $E_1$, the image $t(\phi)$ $\in$ $E_2$ is satisfiable if and only if $\phi$ is satisfiable. The function $t$ is said to \textit{preserve satisfiability}.
\end{flushleft}
\begin{flushleft}
\textbf{Definition 11}. Let $E_1$ and $E_2$ be sets of Boolean formulas. A mapping from $E_1$ to $E_2$ that \textit{preserves satisfiability over conjunction} is a function  $t:E_1$ $\rightarrow$ $E_2$ such that, for every pair of formulas $\phi_1,$ $\phi_2$ $\in$ $E_1$, the formula $t(\phi_1)$ $\wedge$ $t(\phi_2)$ is satisfiable if and only if $\phi_1$ $\wedge$ $\phi_2$ is satisfiable. The function $t$ is said to be \textit{satisfiability-preserving over conjunction}.
\end{flushleft}

\begin{flushleft}
\textbf{Definition 12}. A set $E$ of Boolean formulas is said to be a \textit{satisfiability representation} of the set $B^n$ $\rightarrow$ $B$ if and only if there exist a full representation $E_1$ of the set $B^n$ $\rightarrow$ $B$ and a function $t:E_1$ $\rightarrow$ $E$ that preserves satisfiability over conjunction. The set $E$ is  said to \textit{satisfiability-represent} the set $B^n$ $\rightarrow$ $B$.
\end{flushleft}

\begin{flushleft}
\textbf{Theorem 2}. Let $E$ be a satisfiability representation of $B^n$ $\rightarrow$ $B$. There \textbf{does not exist} a Post machine M such that, for every pair of formulas $\phi_1,$ $\phi_2$ $\in$ $E$, M$(\phi_1,$ $\phi_2)$ correctly decides  whether the formula $\phi_1$ $\wedge$ $\phi_2$ is satisfiable by following a terminated path that includes fewer than $2^n$ conditional branches.
\end{flushleft}

\begin{flushleft}
\textbf{Proof}. Our proof for Theorem 2 is essentially identical to that for Theorem 1, with the following adaptions:
\end{flushleft}
\begin{itemize} [align=left,leftmargin=*]
\item[1.] The proof assumes that  there exists a Post machine M such that, for every pair of formulas $\phi_1,$ $\phi_2$ $\in$ $E$, M$(\phi_1,$ $\phi_2)$ correctly decides  the satisfiability of  $\phi_1$ $\wedge$ $\phi_2$ by following a terminated path that includes fewer than $2^n$ conditional branches. 

\item[2.] Since $E$ satisfiability-represents $B^n$ $\rightarrow$ $B$, there is a set $E_1$ that is a full representation of $B^n$ $\rightarrow$ $B$ and there is a function  $t:E_1$ $\rightarrow$ $E$ that preserves satisfiability over conjunction. Let the set S contain, for each function $f:B^n$ $\rightarrow$ $B$, one run of M with $(t(\phi_f),$ $t(\phi_{\overline{f}}))$ as its bipartite input, where $\phi_f,$ $\phi_{\overline{f}}$ $\in$ $E_1$ and, hence, $t(\phi_f),$ $t(\phi_{\overline{f}})$ $\in$ $E$.

\item[3.] For every function $f:B^n$ $\rightarrow$ $B$ and for every assignment, one of $f$ and $\overline{f}$ evaluates to $false$. Since $\phi_f$ and $\phi_{\overline{f}}$ represent $f$ and $\overline{f}$,  for every assignment one of $\phi_f$ and $\phi_{\overline{f}}$ evaluates to $false$. Hence, $\phi_f$ $\wedge$ $\phi_{\overline{f}}$ is $false$ for all assignments and, thus, is not satisfiable. Since $t$ is satisfiability-preserving over conjunction,  the formula  $t(\phi_f)$ $\wedge$ $t(\phi_{\overline{f}})$ is not satisfiable. Hence, every run in S must eventually reject its input.

\item[4.] To derive a contradiction, let M$(t(\phi_g),$ $t(\phi_{\overline{g}}))$ and  M$(t(\phi_h),$ $t(\phi_{\overline{h}}))$ be two runs in S that reject their inputs by following a common terminated path P that includes fewer than $2^n$ conditional branches - as deptailed in the proof for Theorem 1, there must be at least two such runs in S. By Lemma 2,  the two runs M$(t(\phi_g),$ $t(\phi_{\overline{h}}))$ and M$(t(\phi_h),$ $t(\phi_{\overline{g}}))$, which are not in S, must follow the same execution path P to  reject their inputs, as the two runs  M$(t(\phi_g),$ $t(\phi_{\overline{g}}))$ and  M$(t(\phi_h),$ $t(\phi_{\overline{h}}))$ do. Since $g$ and $h$ are different Boolean functions, there exists an assignment $s$ that makes $g$ and $h$ evaluate to different values. Therefore, $g$ and $\overline{h}$ evaluate to the same value on the assignment $s$. If both $g$ and $\overline{h}$ evaluate to $true$ on the assignment $s$, then so will both of the formulas $\phi_g$ and $\phi_{\overline{h}}$ since $\phi_g$ and $\phi_{\overline{h}}$ represent $g$ and $\overline{h}$. Hence, $\phi_g$ $\wedge$ $\phi_{\overline{h}}$ is satisfiable.  Since $t$ is satisfiability-preserving over conjunction, $t(\phi_g)$ $\wedge$ $t(\phi_{\overline{h}})$ is satisfiable too. On the other hand, if both $g$ and $\overline{h}$  evaluate to $false$ on the assignment $s$, then both $h$ and $\overline{g}$ evaluate to $true$ on the assignment $s$, and the formula $t(\phi_h)$ $\wedge$ $t(\phi_{\overline{g}})$ can be similarly shown to be satisfiable. So, at least one of the formulas $t(\phi_g)$ $\wedge$ $t(\phi_{\overline{h}})$ and $t(\phi_h)$ $\wedge$ $t(\phi_{\overline{g}}))$ is satisfiable. That is, at least one of the two runs M$(t(\phi_g),$ $t(\phi_{\overline{h}}))$ and M$(t(\phi_h),$ $t(\phi_{\overline{g}}))$ \textbf{should accept} its input. However, as discussed previously, by Lemma 2 both M$(t(\phi_g),$ $t(\phi_{\overline{h}}))$ and M$(t(\phi_h),$ $t(\phi_{\overline{g}}))$ \textbf{reject} their inputs. That is, by Lemma 2, at least one  of the two runs M$(t(\phi_g),$ $t(\phi_{\overline{h}}))$ and M$(t(\phi_h),$ $t(\phi_{\overline{g}}))$ \textbf{incorrectly} rejects its input. This contradicts the assumption stated above in item 1. \textbf{Q.E.D.}
\end{itemize}

We give an example of a set of restricted Boolean formulas that satisfiability-represents $B^n$ $\rightarrow$ $B$.  It is well known that the problem of CNF satisfiability can be reduced to 3CNF satisfiability, e.g., [2,5]. Specifically, when this reduction is applied to a CNF formula $C_1$ $\wedge$ $C_2$, where   $C_1$ and $C_2$ are CNF formulas, the  reduction yields a formula $t(C_1)$ $\wedge$ $t(C_2)$ as the resultant 3CNF formula, where $t(C_1)$ and $t(C_2)$ are 3CNF formulas and are derived by applying the reduction to $C_1$ and $C_2$ respectively. The formulas $t(C_1)$ and $t(C_2)$ are satisfiable if and only if $C_1$ and $C_2$ are, respectively, and the resultant 3CNF formula $t(C_1)$ $\wedge$ $t(C_2)$ is satisfiable if and only if the original CNF formula $C_1$ $\wedge$ $C_2$ is. This reduction  introduces distinct new variables into the resultant 3CNF formulas. With the new variables being distinct, this reduction defines a mapping from CNF formulas to 3CNF formulas that is satisfiability-preserving over conjunction. Let $E_1$ be a set of CNF formulas that fully represents $B^n$ $\rightarrow$ $B$. This reduction can be used to transform each CNF formula in $E_1$ into a 3CNF formula. Let $E$ be the set of the resultant 3CNF formulas. The set $E$ satisfiability-represents the set $B^n$ $\rightarrow$ $B$.

The following corollary directly follows from Theorem 2.

\begin{flushleft}
\textbf{Corollary 2.1}. Let $E$ be a set of 3CNF formulas that satisfiability-represents $B^n$ $\rightarrow$ $B$. There \textbf{does not exist} a Post machine M such that, for every pair of 3CNF formulas $\phi_1,$ $\phi_2$ $\in$ $E$, M$(\phi_1,$ $\phi_2)$ correctly decides  whether the 3CNF formula $\phi_1$ $\wedge$ $\phi_2$ is satisfiable by following a terminated path that includes fewer than $2^n$ conditional branches. 
\end{flushleft}

Similarly, there is a reduction from the problem of DNF falsifiability to 3DNF falsifiability [1].  By duality, the following corollary follows from Corollary 2.1.  The term \textit{falsifiability-represent} is the dual of the term satisfiability-represent defined previously. A detailed definition of the term falsifiability-represent parallels Definitions 11-12.

\begin{flushleft}
\textbf{Corollary 2.2}. Let $E$ be a set of 3DNF formulas that falsifiability-represents $B^n$ $\rightarrow$ $B$. There \textbf{does not exist} a Post machine M such that, for every pair of 3DNF formulas $\phi_1,$ $\phi_2$ $\in$ $E$, M$(\phi_1,$ $\phi_2)$ correctly decides  whether the 3DNF formula $\phi_1$ $\vee$ $\phi_2$ is falsifiable by following a terminated path that includes fewer than $2^n$ conditional branches.
\end{flushleft}

\section{Discussion}
\subsection{Turing Machines}

Obviously a Post machine is similar to a Turing machine [6] with a two-way tape of which each "square" can be either blank or marked. A lower bound of $2^n$ binary decisions on such a Turing machine can be established by proofs that are similar to those provided in this paper. 

A difference between the two models is that a Turing machine bundles each decision with a "move", which includes a state transition, a movement of the read/write head, and an operation to write a symbol to a tape square. Hence, the lower bound $2^n$ is on the number of moves that a Turing machine makes, whereas in the context of a Post machine, the lower bound is on the number of conditional branches executed and does not apply to other types of instructions that a Post machine executes.

Another difference is that, in general, a Turing machine has a finite tape alphabet which may consist of more than two symbols. If a Turing machine has $k$ symbols in its tape alphabet, then, for each move, the machine makes a $k$-way decision, instead of a two-way decision. By proofs [3] similar to those given in the previous sections, it can be shown that a lower bound on the number of moves that a Turing machine with $k$ symbols in its tape alphabet needs to make in order to decide Boolean satisfiability is $2^nlog_k2$.

\subsection{Non-Sequential Access and Number of Partitions}

The lower bound established in this paper does not depend on sequentiality of access to the boxes in the symbol space. Hence, the lower bound will hold even if a Post machine is capable of non-sequential access to the boxes. For example, even if an instruction is allowed to specify that the head move to a box at a specific address or that the head skip a certain number of boxes to the left or to the right, the lower bound of $2^n$ conditional branches will still hold. The proofs only require straightforward adaptions to accommodate this flexibility.

Additionally, the lower bound will still hold if a symbol space convention divides the symbol space into more than two disjoint partitions, although only two partitions are used to hold a bipartite input. The proofs only require minor adaptions to accommodate this flexibility. 
\begin{flushleft}
\textbf{References}
\end{flushleft}
\begin{itemize}[align=left,leftmargin=*]
\item[1.] Cook, S.A. The complexity of theorem proving procedures. In \textit{Proceedings, Third Annual ACM Symposium on the Theory of Computing} (1971), pp. 151-158.
\item[2.]Hopcroft, J.E., Ullman, J.D., \textit{Introduction to Automata Theory, Languages, and Computation}. Addison-Wesley, 1979.
\item[3.] Hsieh, S. C. A Lower Bound for Boolean Satisfiability on Turing Machines. preprint (2014) available at \textit{http://arxiv.org/abs/1406.5970} 
\item[4.] Post, E.L. Finite Combinatory Processes-Formulation 1. \textit{The Journal of Symbolic Logic} 1 (1936) pp. 103-105.
\item[5.] Sipser, M. \textit{Introduction to the Theory of Computation}. 2nd ed. Thomson Course Technology, 2006.
\item[6.] Turing, A.M. On Computable Numbers, with an Application to the Entscheidungs problem. In \textit{Proceedings of the London Mathematical Society} (1936), pp.230-265.
\end{itemize}

\end{document}